\documentclass[12pt]{article}
\usepackage[letterpaper,hmargin=25mm,vmargin=25mm,nohead]{geometry}

\usepackage{bm}
\newcommand* {\vek}[1]{\ensuremath{\bm{\mathrm{#1}}}}
\newcommand* {\vekc}[1]{\ensuremath{\bm{\mathcal{#1}}}}
\newcommand* {\kk}{\vek{k}}
\newcommand* {\kbolz}{k_\mathrm{B}}
\newcommand* {\bohrmag}{\mu_\mathrm{B}}
\newcommand* {\fermi}{\mathrm{F}}
\newcommand* {\eps}{\ensuremath{\varepsilon}}
\newcommand* {\crossref}{\emph{[elsewhere]}}

\usepackage{amssymb}
\usepackage{graphicx}
\usepackage{natbib}
\usepackage{titlesec}
\titleformat{\section}
 {\normalfont\large\bfseries}{\thesection}{1em}{}
\titlespacing*{\section}{0pt}{*3}{*1}
\titleformat{\subsection}
 {\normalfont\normalsize\bfseries}{\thesubsection}{1em}{}
\titlespacing*{\subsection}{0pt}{*2}{*1}

\begin{document}
\begin{center}
{\large\bf Spin-dependent transport of carriers in semiconductors}\\[3ex]
R.~Winkler\\[2ex]
Department of Physics, Northern Illinois University,
DeKalb, IL 60115, USA
\\[2ex]\today
\end{center}


\begin{quote}
  This article reviews spin-dependent transport of carriers in
  homogenous three-dimensional and two-dimensional semiconductors.
  We begin with a discussion of optical orientation of electron
  spins, which allows both the creation and detection of
  spin-polarized carriers in semiconductors. Then we review
  non-equilibrium spin flow including spin drift and diffusion
  caused by electric fields and concentration gradients. A
  controlled spin precession is possible both in external magnetic
  fields and in effective magnetic fields due to a broken inversion
  symmetry. Although the Coulomb interaction does not couple to the
  spin degree of freedom, it affects the spin-dependent transport
  via the spin Coulomb drag. In gyrotropic media, the optical
  creation of spin-oriented electrons gives rise to spin
  photocurrents, which reverse their direction when the radiation
  helicity is changed from left-handed to right-handed. The reverse
  process is possible, too, i.e., an electric current in a
  gyrotropic medium gives rise to a spin polarization in the bulk of
  the sample.
\end{quote}
\section{Introduction}

Broadly speaking, spin-dependent transport phenomena in
semiconductors can be divided into two categories. On the one hand,
we have those effects such as spin drift, spin diffusion, and spin
precession that refer to the transport of spin-polarized carriers.
These effects are of central importance for spintronics device
concepts where the generation of spin-polarized distributions of
carriers are spatially separated from those elements that manipulate
and detect the spins. On the other hand, we also have spin-dependent
phenomena such as the spin Coulomb blockade or weak localization and
spin-split Shubnikov-de Haas oscillations visible in
magneto-transport of two-dimensional (2D) electron systems. These
effects provide important insights into the nature of the
spin-dependent interactions, such as exchange and spin-orbit
coupling, that can be exploited for the manipulation of spin
systems. In this review, we will focus mostly on the former class of
phenomena. Also, we will focus mostly on homogenous systems and
touch only briefly on the properties of structured devices that are
discussed \crossref.

We begin with a discussion of optical orientation of electron spins
in semiconductors (Section~\ref{sec:optor}). Then we review
non-equilibrium spin flow including spin drift and diffusion
(Section~\ref{sec:drift}), and spin precession
(Section~\ref{sec:prec}). Coulomb effects in spin transport are
discussed in Section~\ref{sec:coul}. Finally, we review in
Section~\ref{sec:charge} spin photocurrents and the reverse effect,
the electrical generation of a spin polarization.

\section{Optical orientation of electron spins}
\label{sec:optor}

Various schemes have been developed to generate spin-polarized
carrier distributions in nonmagnetic semiconductors. Broadly
speaking, these fall into three categories. First, optical
excitation allows creation of spin-polarized electrons inside the
semiconductor. Second, magnetic layers can be used to inject
spin-polarized carriers into the semiconductors. These magnetic
layers can be either ferromagnetic or semimagnetic semiconductors
(see \crossref) or ferromagnetic metal electrodes attached to the
semiconductor (see \crossref). Finally, dynamic phenomena based on
electric fields and charge currents can give rise to spin
polarization inside the semiconductor or spin accumulation at the
edges of the sample (see Section~\ref{sec:elspingen} and \crossref).

\begin{figure}[t]
\includegraphics[width=0.48\linewidth]{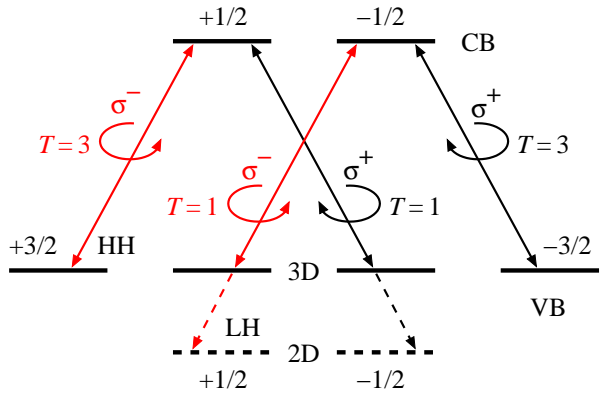}
\hfill
\raisebox{0ex}{\begin{minipage}[b]{0.48\linewidth}
  \caption{\label{fig:selectrule}Selection rules and relative
  transition rates $T$ for optical transitions between valence band
  (VB) states having an effective spin $S=3/2$ and conduction band
  (CB) states with $S=1/2$ \citep{dya84}). In bulk semiconductors
  the HH states ($S_z = \pm 3/2$) are degenerate with the LH states
  ($S_z = \pm 1/2$) whereas in quasi-2D systems the LH states
  (dashed bold lines) are lower in energy than the HH states.}
\end{minipage}}
\end{figure}

Here we will focus on the \emph{optical orientation} of electrons
that has proven to be a powerful technique since some of the
earliest studies of spin-related phenomena in semiconductors were
performed \citep{lam68, mei84}. In direct semiconductors like GaAs,
the electron states in the conduction band have spin $S=1/2$,
whereas the hole states in the valence band have an effective spin
$S=3/2$. The hole states with spin $z$ component $S_z = \pm 3/2$ are
denoted heavy-hole (HH) states, whereas the light-hole (LH) states
have $S_z = \pm 1/2$. Left (right) circularly polarized photons
carry a $z$ component of angular momentum of $-1$ ($+1$) so that
conservation of angular momentum results in the selection rules for
optical transitions depicted in Figure~\ref{fig:selectrule}. A more
detailed analysis shows that the probabilities for transitions from
the HH states to the conduction band are three times larger than the
probability for optical transitions from the LH states. In bulk
semiconductors, the maximum attainable degree of spin polarization
is thus $P=50$\%, where $P$ is defined as
\begin{equation}
\label{eq:PolgradAllgemein}
P = \frac{N_+ - N_-}{N_+ + N_+} \, ,
\end{equation}
and $N_+$ ($N_-$) is the number of electrons with spin up (down),
respectively. In 2D systems the degeneracy of the HH and LH states
is lifted as sketched in Figure~\ref{fig:selectrule}. For resonant
excitation at the HH energy we thus expect a rise of the maximum
attainable degree of polarization up to $P = 100$\%.

A particular advantage of the optical orientation scheme lies in the
fact that it holds both for absorption and emission, so that it can
be used for creating and for detecting spin-polarized carrier
distributions. However, the holes lose their spin orientation
significantly faster than the electrons, and the oriented electrons
can recombine with \emph{any} hole. Therefore,
Figure~\ref{fig:selectrule} implies that the polarization of the
recombination photoluminescence (PL) in bulk semiconductors does not
exceed $\sim 25$\%. (This does not apply to 2D systems where the
recombination PL is due to a transition from the lowest electron to
the lowest HH state.)

Even in a single-particle picture for the optical excitation, the
$3:1$ ratio of HH and LH transitions is obtained only if HH-LH
coupling of the hole states at nonzero wave vectors $\bm{k}$ is
neglected. Due to this HH-LH coupling, the hole states with $k>0$
are not spin eigenstates. Furthermore, a realistic picture must take
into account that optical absorption gives rise to the formation of
excitons, i.e., Coulomb correlated electron-hole pairs. Thus even
for excitations close to the absorption edge we get substantial
HH-LH coupling because the exciton states consist of electron and
hole states with $k$ of the order of $1/a_\mathrm{B}^\ast$, where
$a_\mathrm{B}^\ast$ is the effective Bohr radius. The Coulomb
coupling between different electron and hole states yields a second
contribution to the mixing of single-particle states with different
values of $S_z$. Finally, we must keep in mind that for higher
excitation energies we get a superposition of exciton continua that
are predominantly HH- or LH-like. These different excitons
contribute oppositely to the spin orientation of electrons. We note
that these arguments are valid for the optical excitation of bulk
semiconductors and quasi-2D systems \citep{pfa05}.

Optical orientation in bulk systems was reviewed by \cite{dya84}.
Early works on 2D systems were published by \citet{wei81a} and
\citet{mas84} who reported on polarization-resolved transmission and
PL experiments on GaAs/AlGaAs quantum wells (QWs) under cw
excitation. In later works, the electron spin polarization in
quasi-2D systems was studied using time-resolved photoluminescence
excitation spectroscopy. For excitation energies even slightly above
the HH resonance, several authors \citep{fre90, dar93a, mun95}
observed a polarization that was significantly smaller than one.
These measurements were carried out on fairly narrow GaAs/AlGaAs
QWs. A first well-width dependent study of optical orientation was
performed experimentally by \citet{rou92}. For energies near the HH
resonance, they found initial spin polarizations in the range of
$60-80$\%.

\begin{figure}[t]
\includegraphics[width=0.4\linewidth]{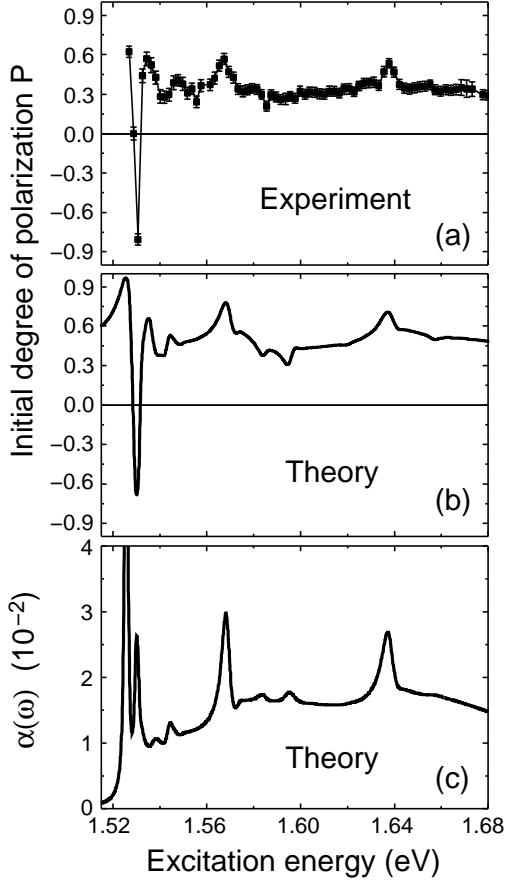}
\hfill
\raisebox{6ex}{\begin{minipage}[b]{0.5\linewidth}
  \caption{\label{fig:198}(a) Measured and (b) calculated degree of
  spin polarization $P$ and (c) calculated absorption coefficient
  $\alpha(\omega)$ as a function of excitation energy of a
  $198$-{\AA}-wide GaAs/AlAs QW. The lowest peak in the absorption
  spectrum is an HH exciton that gives rise to a large positive spin
  orientation. The next peak slightly above the first peak is due to
  the LH exciton, and it results in a strongly negative spin
  orientation. The peaks at higher energies are Fano resonances,
  which give mixed contributions to the spin polarization. [Adapted
  from \citet{pfa05}.]}
\end{minipage}}
\end{figure}

\citet{twa87} as well as \citet{uen90a} studied the polarization of
QW PL theoretically, taking into account HH-LH coupling in the
valence band. However, these authors neglected the Coulomb
interaction between electron and hole states. On the other hand,
\citet{mai93} investigated the spin dynamics of excitons taking into
account the exchange coupling between electrons and holes, but
disregarded the HH-LH coupling in the valence band. Recently, a
detailed experimental and theoretical study of optical orientation
in 2D systems was performed by \citet{pfa05} confirming that the
polarization of the measured PL reflects the spin polarization of
the excited electrons [equation (\ref{eq:PolgradAllgemein})] over a
wide range of excitation energies. As an example,
Figure~\ref{fig:198} shows the measured and calculated degree of
spin polarization $P$ as a function of excitation energy of a
$198$-{\AA}-wide GaAs/AlAs QW.

\section{Non-equilibrium spin flow in semiconductors}
\label{sec:spinflow}

\subsection{Spin drift and diffusion}
\label{sec:drift}

Similar to electric charge distributions in semiconductors, a
non-equilibrium spin distribution can spread out diffusively or it
can drift in the presence of an electric field. However, these
phenomena behave qualitatively different in $p$- and $n$-type
semiconductors \citep{dya71a}. In $p$-type semiconductors, only the
spins of the non-equilibrium electrons become oriented. Their number
is proportional to the intensity of the light, but the degree of
orientation does not depend on the intensity
(Figure~\ref{fig:selectrule}). As drift and diffusion of the spin
orientation must preserve charge neutrality, the kinetics of the
spin orientation follows the kinetics of the charge distribution.
Charge diffusion in doped semiconductors is characterized by the
diffusion coefficient of the minority carriers \citep{smi78}.
Therefore, electron spin diffusion in $p$-type semiconductors is
essentially characterized by the charge diffusion coefficient for
electrons.%
\footnote{When a semiconductor has a large absorption coefficient
near the band edge, an emitted photon is usually reabsorbed before
it can escape the crystal \citep{dum57}. The detailed analysis of
spin diffusion in $p$-type semiconductors performed by \citet{gar76}
and \citet{gio94} showed that allowance for diffusion and
reabsorption was essential for the proper interpretation of their
experimental data. Even in $n$-type GaAs it was found that
reabsorption can be important for spin diffusion \citep{dzh97}.
Please note that the first authors of the latter two publications
are, in fact, the same.}

In $n$-type semiconductors the situation is qualitatively different
due to the fact that the optically excited electrons augment the
equilibrium electrons \citep{dya71a}. Therefore, significant optical
orientation of electron spins is possible at moderate degrees of
excitation when the excess photoelectron density is still much less
than the equilibrium electron density. The mechanism underlying this
effect is the following. Absorption of circularly polarized light
creates electrons with mainly a single spin orientation. The spin
relaxation time~$\tau_s$ of these electrons is typically much
greater than the excess carrier lifetime. Holes, on the other hand,
have a short spin relaxation time so that the spin orientation of
the optically created holes is quickly lost. Therefore \emph{any}
electron can recombine with these holes with a recombination rate
that is usually independent of the sign of the spin. Thus, optical
excitation is a source for spin-polarized electrons whereas
recombination represents a drain for electrons with the ``wrong''
spin orientation. Under stationary excitation, the oriented
electrons are the equilibrium ones.

In a bulk sample, the light is usually absorbed in a narrow layer
near the surface of the crystal. In this case the excess carriers
penetrate a distance of the order of the diffusion length $L =
\sqrt{D_p \,\tau}$, where $D_p$ is the hole diffusion coefficient in
the $n$-type sample (which is usually small), and $\tau$ is the
lifetime of the non-equilibrium carriers. On the other hand, the
orientation penetrates a depth of the order of the spin diffusion
length $L_s = \sqrt{D_s \,\tau_s}$, where $D_s$ is the spin
diffusion coefficient of the electrons (which is usually large; it
is approximately equal to the electron diffusion coefficient $D_e$).
Under typical experimental conditions we thus have $L_s \gg L$ in an
$n$-type sample \citep{dya71a}. Beyond a layer of thickness $\sim
L$, recombination cannot change the degree of polarization $P$ that
falls off like $P(z) = P(0) \exp(-z/L_s)$, i.e., a spin orientation
of the order of $P(0)$ penetrates into a layer of depth $\sim L_s$,
where there are no excess carriers. [A small number of photoexcited
carriers can be present within this layer because of reabsorption
\citep{dzh97}.]

In general, the motion of the spin density $\vek{S}$ is
characterized by a drift-diffusion equation \citep{dya76, gar76,
dya84}
\begin{equation}
  \label{eq:drift}
  \frac{\partial \vek{S}}{\partial t}
  = D_s \, \nabla^2 \vek{S}
  + \frac{e\vekc{E}\cdot\nabla \,\vek{S}}{\kbolz T}
  + \vek{\Omega} \times \vek{S}
  - \frac{\vek{S}}{\tau_s}
  - \frac{\vek{S} - \vek{S}_0}{\tau},
\end{equation}
similar to the drift and diffusion of charge. Here $\vekc{E}$ is a
built-in or external electric field; $T$ is the temperature; and
$\Omega$ is the spin precession frequency, which can be due to an
external magnetic field $\vek{B}$, i.e., $\vek{\Omega} = g^\ast
\bohrmag \vek{B}/\hbar$, or due to spin-orbit coupling at $B=0$, see
Section~\ref{sec:prec} below. The last two terms reflect two reasons
for the spin orientation to vanish, spin relaxation and
recombination, where $\vek{S}_0$ is the average spin orientation at
the moment of photocreation. Recently, the drift-diffusion equation
(\ref{eq:drift}) was reconsidered by \citet{fla00} and \citet{yu02}.

Spin drift and diffusion have been studied experimentally by several
groups. \citet{dzh97} estimated that the spin diffusion length in
their $n$-type GaAs sample was $L_s = 10~\mu$m. \citet{hae98} found
that the spin orientation in intrinsic GaAs was almost completely
preserved over a distance of $4~\mu$m. \citet{kik99} performed a
detailed study of spin transport in intrinsic and $n$-type GaAs
samples in which gates allowed one to stir the drift of the
spin-polarized electrons. Using non-local Faraday rotation, they
found that the drift distance of the spin-oriented electrons was
linear in the electric field, and it could exceed a distance of
$100~\mu$m for electric fields of 16~V~cm$^{-1}$. \citet{fie99} used
semimagnetic Be$_x$Mn$_y$Zn$_{1-x-y}$Se to inject spin-polarized
electrons into a $0.1$-$\mu$m-thick layer of $n$-type AlGaAs
followed by a 15-nm-wide GaAs, where the spin-polarized electrons
recombined with holes that were injected from the other side of the
QW (a spin light-emitting diode). In a similar experiment,
\citet{ohn99a} used ferromagnetic GaMnAs to inject spin-polarized
electrons into an intrinsic layer of GaAs, followed by an InGaAs QW.

It has been found that interfaces between different semiconductors
do not affect spin transport. This was first noticed by
\citet{gar76}, who studied spin orientation for a sample that
contained a GaAs QW embedded in thick graded layers of $p$-type
Al$_x$Ga$_{1-x}$As. \citet{mal00} found that even the interface
between ZnSe and GaAs, a II-VI and a III-V semiconductor, did not
suppress spin transport.

Lateral spin diffusion was studied by \citet{cam96}. When two laser
beams with crossed polarizations interfere, the light intensity on
the sample is uniform, but the polarization alternates between left
polarized, linear, and right polarized. Therefore, a spin grating is
generated in the sample where the optical orientation of the
electrons alternates across the excitation region. By analyzing the
orientation decay as a function of time, these authors could
determine the spin diffusion coefficient $D_s$ and the spin
relaxation time $\tau_s$. The spin diffusion length $L_s = \sqrt{D_s
\,\tau_s}$ appeared to be approximately $8~\mu$m \citep{kav02}.

\subsection{Spin precession}
\label{sec:prec}

The magnetic-field-dependent term $\vek{\Omega} \times \vek{S}$ in
equation (\ref{eq:drift}) describes the precessional motion of the
oriented spins in an external field $\vek{B}$ or an effective field
due to spin-orbit coupling. For a transverse external field
$\vek{B}$, it gives rise to the Hanle effect, a depolarization of
luminescence induced by the field $\vek{B}$ \citep{dya84}. In a
homogenous system [i.e., $\nabla \,\vek{S} = 0$ in equation
(\ref{eq:drift})], we get the expression for the Hanle curve
\begin{equation}
  \label{eq:hanle}
  S_z (B) = \frac{S_z(0)}{1 + (\Omega T_s)^2},
  \hspace{3em}\mbox{where}\hspace{2em}
  S_z (0) = \frac{S_0}{1 + \tau/\tau_s}.
\end{equation}
Here we have assumed that the $z$ direction is the direction of the
exciting radiation with $\vek{S}_0 \perp \vek{B}$, and $T_s$ is the
``spin lifetime'' defined by $T_s^{-1} = \tau^{-1} + \tau_s^{-1}$.
From the Hanle curve as a function of field $B$, one can thus
extract the lifetime $\tau$ and the spin relaxation time $\tau_s$ of
the carriers (provided the effective Land\'e factor $g^\ast$ is
known). However, a particular situation arises in $n$-type
semiconductors where recombination is not possible past the surface
layer of thickness $\sim L$. The depolarization induced by the
magnetic field thus changes the gradient of the degree of
polarization within the layer where electrons are oriented.
Therefore, the spin diffusion rate becomes magnetic-field-dependent,
which results in a distinct change of the functional form of the
Hanle curve as a function of magnetic field \citep{dya76}.

\begin{figure}[tbp]
  \includegraphics[width=0.6\linewidth]{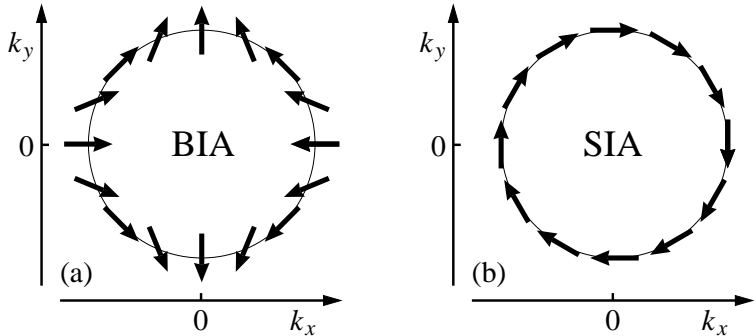}
  \hfill
  \raisebox{3ex}{\begin{minipage}[b]{0.35\linewidth}
    \caption[]{\label{fig:sbia_001}Effective magnetic field
    $\vek{\Omega}(\kk_\|)$ along a contour of constant energy (a) due
    to the Dresselhaus term in a system with BIA and (b) due to the
    Rashba term in a system with SIA.}
  \end{minipage}}
\end{figure}

In the presence of both time-inversion symmetry and space-inversion
symmetry, all electron states in a solid with a given wave vector
$\kk$ are twofold degenerate. When the potential through which the
carriers move is inversion-asymmetric, however, the spin degeneracy
is removed even in the absence of an external magnetic field $B$. We
then obtain two branches of the energy dispersion, $E_+(\kk)$ and
$E_-(\kk)$. This spin splitting can be the consequence of a bulk
inversion asymmetry (BIA) of the underlying crystal [e.g., a zinc
blende structure \citep{dre55a}], and of a structure inversion
asymmetry (SIA) of the confinement potential \citep{ohk74, byc84}.
Strain gives rise to a third contribution to $B=0$ spin splitting
\citep{sei77, how77}. A fourth contribution can be the low
microscopic symmetry of the atoms at an interface~\citep{ros02}.
$B=0$ spin splitting has been reviewed, e.g., by \citet{pik88} and
\citet{win03}. In the present context it is important that the spin
splitting can be ascribed to an effective Zeeman term $H = (\hbar/2)
\: \vekc{\sigma} \cdot \vek{\Omega}(\kk)$ with an effective magnetic
field $\vek{\Omega}(\kk)$. In leading order of $\kk$, the effective
field in a 2D electron system on a (001) surface reads
\begin{equation}
  \label{eq:omega_2d}
\vek{\Omega}(\vek{k}_\|)
=
\frac{2\gamma}{\hbar} \left(\begin{array}{c}
    k_x \left(k_y^2 - \langle k_z^2 \rangle\right) \\[0.5ex]
    k_y \left(\langle k_z^2 \rangle - k_x^2\right) \\[0.5ex]
    0
    \end{array} \right) +
\frac{2\alpha}{\hbar} \left(\begin{array}{c}
    k_y \\[0.5ex]
    -k_x \\[0.5ex]
    0
  \end{array} \right).
\end{equation}
The first term characterizes the BIA spin splitting of the electron
states. It is called the Dresselhaus or $k^3$ term \citep{dre55a,
bra85}. It exists already in bulk zinc blende semiconductors due to
the broken inversion symmetry. In quasi-2D systems only the in-plane
wave vector $\vek{k}_\| = (k_x, k_y, 0)$ is a continuous variable.
In first-order perturbation theory, the wave vector components $k_z$
and powers thereof are replaced by expectation values with respect
to the subband wave functions. The field $\vek{\Omega}(\vek{k}_\|)$
due to BIA is depicted in Figure~\ref{fig:sbia_001}(a). We note that
in 2D systems, the Dresselhaus term depends on the crystallographic
orientation of the substrate. For 2D systems on an $[mmn]$ surface
with integers $m$ and $n$, the Dresselhaus term was given by
\citet{bra85}.

The momentum scattering of electrons off other electrons, impurities,
phonons, etc., results in a random walk of oriented electrons in the
field $\vek{\Omega}(\vek{k})$, which gives rise to the so-called
Dyakonov-Perel spin relaxation \citep{dya71} discussed \crossref.
A \emph{controlled} precession of electrons in the Dresselhaus field
$\vek{\Omega}(\vek{k})$ was first demonstrated by \citet{rie84}, who
investigated the polarization of photoemission following optical
orientation. After deposition of Cs and O on the (110) surface of
their strongly $p$-doped GaAs sample, a surface inversion layer was
formed where the bands were strongly bent downward. Electrons moving
through this layer gain a large kinetic energy. Yet if the layer is
very narrow, they move ballistically with $\kk$ normal to the
surface so that the direction of $\vek{\Omega}$ is the same for all
escaping electrons. The photoelectron orientation is thus rotated
away from the initial direction defined by the pumping light beam,
as observed by \citeauthor{rie84}%
\footnote{The Rashba spin splitting discussed below was not taken
into account by \citet{rie84} for the interpretation of their
experiment. However, this does not change the qualitative picture.}

In asymmetric QWs, SIA gives rise to the second term in equation
(\ref{eq:omega_2d}), which is frequently called the Rashba term
\citep{ras60, byc84}. The coefficients $\gamma$ and $\alpha$ depend
on the underlying semiconductor bulk material. But $\alpha$ depends
also on the asymmetry of the QW in the growth direction. It can be
tuned by means of front and back gates \citep{nit97}. This is
exploited in the famous spin field-effect transistor proposed by
\citet{dat90}, which is discussed \crossref. The field
$\vek{\Omega}(\vek{k}_\|)$ due to SIA is depicted in
Figure~\ref{fig:sbia_001}(b).

A third contribution to $\vek{\Omega} (\kk)$ at $B=0$ is obtained by
means of strain. In lowest order of $\kk$ and of the components
$\eps_{ij}$ of the strain tensor we get \citep{pik84}
\begin{equation}
  \label{eq:omega_strain}
\vek{\Omega}(\vek{k})
=
\frac{C_3}{\hbar} \left(\begin{array}{c}
    \eps_{xy} k_y - \eps_{xz} k_z \\[0.5ex]
    \eps_{yz} k_z - \eps_{yx} k_x \\[0.5ex]
    \eps_{zx} k_x - \eps_{zy} k_y
    \end{array} \right)
+ \frac{C_3'}{\hbar} \left(\begin{array}{c}
    k_x (\eps_{yy} - \eps_{zz}) \\[0.5ex]
    k_y (\eps_{zz} - \eps_{xx}) \\[0.5ex]
    k_z (\eps_{xx} - \eps_{yy})
    \end{array} \right).
\end{equation}
The first term depends on the off-diagonal components of the strain
tensor, i.e., these components describe a shear strain. They are
nonzero, e.g., when uniaxial stress is applied in the
crystallographic directions [111] or [110] of a bulk crystal
\citep{tre79}. The prefactor $C_3'$ of the second term in equation
(\ref{eq:omega_strain}) is nonzero only because of coupling to
remote bands outside the usual $8\times 8$ Kane Hamiltonian.
Therefore, this term is rather small, so usually it can be neglected
\citep{pik84, dya86a}.

For bulk InSb, the effect of strain on spin splitting has been
studied by measuring Shubnikov-de Haas oscillation \citep{sei77} and
cyclotron resonance \citep{ran79}. \citet{dya86a} analyzed the Hanle
effect in the presence of strain in order to obtain $C_3 =
20$~eV{\AA} for GaSb, $C_3 = 5$~eV{\AA} for GaAs, and $C_3 =
3$~eV{\AA} for InP. The decrease of these values from GaSb to InP
correlates with the decrease in the spin-orbit interaction gap in
these crystals from $0.82$~eV to $0.11$~eV.

\begin{figure}[tbp]
  \includegraphics[width=0.48\linewidth]{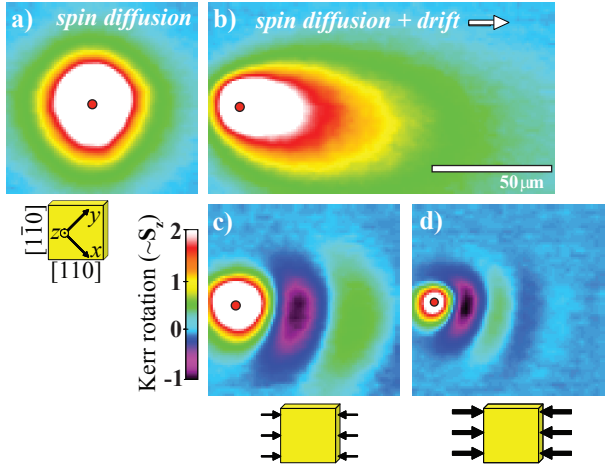}
  \hfill
  \raisebox{1ex}{\begin{minipage}[b]{0.5\linewidth}
    \caption[]{\label{fig:crooker}Images of the electron spin flow
    in a $1$-$\mu$m-thick $n$-GaAs epilayer ($n=1\times
    10^{16}$~cm$^{-3}$) at 4~K, acquired via Kerr-rotation
    microscopy. (a) shows the spin polarization due to spin
    diffusion alone. In (b)-(d), a lateral electric field
    $E=10$~V~cm$^{-1}$ was applied. Finally, $E$ was complemented by
    (c) weak and (d) large uniaxial stress along [110]. The white
    bar in (b) gives the length scale for all four panels. [Adapted
    from \citet{cro05}.]}
  \end{minipage}}
\end{figure}

The effect of strain on spin transport in $n$-type (001) GaAs was
first studied by \citet{kat04a} using time and spatially resolved
Faraday rotation spectroscopy. However, they did not quantify or
tune the strain in their samples. The implications of equation
(\ref{eq:omega_strain}) have been confirmed in detail in experiments
by \citet{cro05}. Similar experiments have been published also by
\cite{bec05}. \citeauthor{cro05} used a small vise to apply
tunable uniaxial strain along the $[110]$ or $[1\overline{1}0]$
direction of their $n$-GaAs sample, while a circularly polarized
$1.58$-eV laser focused to a $4$-$\mu$m spot was used to create
locally a spin orientation along [001]. In spatially resolved
measurements using Kerr rotation they studied the electron spin
precession while the electrons drifted and diffused away from the
position of the laser spot, where the spin orientation was created
(see Figure~\ref{fig:crooker}).

\citeauthor{cro05} found that the spin precession of electrons
drifting and diffusing in the strain field (\ref{eq:omega_strain})
is more robust than the precession of electrons moving in an
external magnetic field. The reason is that in a transverse magnetic
field the ensemble spin orientation dephases quickly when the
precession period falls below the spin diffusion length (the Hanle
effect discussed above). The strain-induced field
(\ref{eq:omega_strain}), on the other hand, is linear in the wave
vector $\kk$ so that slowly moving electrons experience a smaller
field $\vek{\Omega}(\vek{k})$ than the faster electrons. But the
distance the electrons must travel for a complete precession period
is the same for slow and fast electrons so that the electrons remain
in phase [Figures \ref{fig:crooker}(c) and (d)]. This argument also
implies that the precession period should be independent of the
magnitude of the external electric field used to push the electrons,
as confirmed by the experiments of \citeauthor{cro05} and
\citeauthor{bec05}

The strain-induced field (\ref{eq:omega_strain}) has a pronounced
dependence on the wave vector $\kk$. If a uniaxial strain is applied
along the direction $[110]$, we have $\Omega = 0$ for $\kk$ along
[001]. This is analogous to the fact that we have no Dresselhaus
spin splitting in symmetric QWs on a (110) surface for $\kk_\|$
along [001] \citep{win03}. Within the (001) plane the $\kk$
dependence of $\vek{\Omega}$ is the same as for the Rashba term, see
Figure~\ref{fig:sbia_001}(b). If in addition to the strain-induced
field (\ref{eq:omega_strain}) an external magnetic field $\vek{B}$
is applied, the electrons with $\vek{\Omega}(\vek{k})$ approximately
parallel to $\vek{B}$ precess faster than the electrons with
$\vek{\Omega}(\vek{k})$ approximately antiparallel to $\vek{B}$.
This was confirmed by the experiments of \citeauthor{cro05}. To show
this they used the fact that the radial diffusion in a pure
strain-induced field (\ref{eq:omega_strain}) \emph{or} an external
magnetic field $\vek{B}$ is independent of the direction of $\kk$,
which reflects the fact that the magnitude of $\vek{\Omega}$ does
not depend on the direction of the $\kk$ vector of the electrons.
The superposition of both fields, on the other hand, results in an
anisotropic total field $\vek{\Omega}$ the magnitude of which
depends on the direction of $\kk$. This is similar to the fact that,
to lowest (i.e., linear) order in $\kk_\|$, the magnitude of both
the Dresselhaus and Rashba spin splitting in 2D systems are
independent of the direction of $\kk_\|$ (see
Figure~\ref{fig:sbia_001}), yet the superposition of both terms
gives rise to anisotropic spin splitting even in linear order of
$\kk_\|$ \citep{and92}.

The interplay of diffusion, drift in electric fields, and precession
in external magnetic fields was studied theoretically by
\citet{qi03} using a semiclassical Boltzmann equation for the
$2\times 2$ spin density matrix in order to cope with the different
length scales of this problem, such as the diffusion length $L$, the
spin diffusion length $L_s$, and the spin precession length. Spin
diffusion equations for systems with Rashba spin-orbit interaction
in an electric field were studied by \citet{ble06}. Drift and
diffusion were also studied theoretically by \citet{hru06} for an
experimental setup similar to the one used by \cite{cro05} as
described above.

\subsection{Coulomb effects}
\label{sec:coul}

So far we have completely neglected the Coulomb interaction between
the electrons. Although this interaction does not couple to the spin
degree of freedom of the electrons, it has a great influence on
spin-dependent transport. This has long been known in the context of
spin diffusion in spin-polarized liquid $^3$He. \citet{leg68} and
\citet{leg70} have shown that the spin polarization gives rise to a
molecular field, and any given spin will then see (and precess
around) a total field that is the sum of the molecular field and the
external field. This molecular field cannot affect the precession of
the total spin density $\vek{S}$, since it is automatically parallel
to it. However, it produces a torque on the spin current which is
present in the equation of continuity for the latter.
\citeauthor{leg70} showed that, as a result, the equation for
$\vek{S}$ in the hydrodynamic limit no longer has a simple form
similar to equation (\ref{eq:drift}) but he derived a significantly
more complicated hydrodynamic-type spin diffusion equation. More
recently, \citet{tak99} have applied these ideas to the spin
diffusion and drift in 2D electron systems. They solved the quantum
kinetic equation derived from the equation of motion for the
non-equilibrium real-time Green's functions in order to show that
the spin rotation term known for $^3$He is indeed also present in
degenerate 2D electron systems at low temperatures.

\begin{figure}[tbp]
  \includegraphics[width=0.5\linewidth]{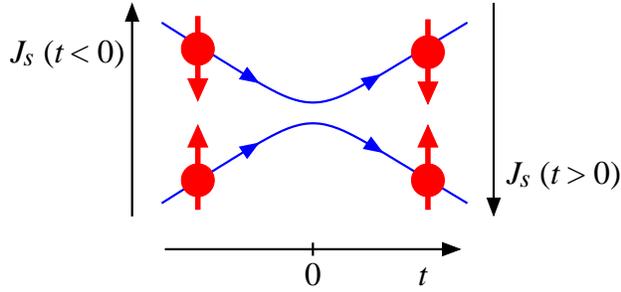}
  \hfill
  \raisebox{1ex}{\begin{minipage}[b]{0.45\linewidth}
    \caption[]{\label{fig:spincurrent}At $t<0$ both electrons
    contribute equally to an upward spin current $\vek{J}_s = (e/m)
    (\vek{P}_\uparrow - \vek{P}_\downarrow)$. At $t=0$, the
    directions of the orbital motions of the electrons are inverted
    due to Coulomb scattering. The direction of the spin current
    $\vek{J}_s$ at $t>0$ is thus reversed.}
  \end{minipage}}
\end{figure}

In a sequence of papers, \citet{dam00, dam01, dam02, dam03}
performed a detailed theoretical analysis of how the Coulomb
interaction affects spin-polarized transport and diffusion in
electron systems [see also \citet{fle01}]. \citeauthor{dam00} showed
that the Coulomb interaction gives rise to a spin Coulomb drag
between the electrons moving with spin up and the electrons moving
with spin down, similar to the Coulomb drag that has been observed
for electrons in two spatially separated layers \citep{gra91,
roj99}. The spin Coulomb drag reflects the fact that while, in the
absence of impurities, the total momentum $\vek{P}=\sum_i \vek{p}_i$
of the electrons is preserved, the ``up'' and ``down'' components of
the total momentum, $\vek{P}_\uparrow = \sum_i
\vek{p}_i(1+\sigma_{zi})/2$ and $\vek{P}_\downarrow = \sum_i
\vek{p}_i(1-\sigma_{zi})/2$, are not separately preserved, even in
the absence of impurities. Here $\vek{p}_i$ is the momentum of the
$i$th electron, and $\sigma_{zi}$ is the Pauli matrix for the $z$
component of the $i$th electron spin. Coulomb scattering can
transfer momentum between spin-up and spin-down electrons, thereby
effectively introducing a ``friction'' for the relative motion of
the two spin components, which tends to equalize the net momenta of
the spin components (see Figure~\ref{fig:spincurrent}).

In a more rigorous formulation, Ohm's law can be written in the form
\begin{equation}
  \newcommand{\arr}[3]{\left(\begin{array}{@{}#1@{}}#2\\[0.6ex]#3
                             \end{array}\right)}
  \label{eq:spindrag}
  \arr{c}{\vekc{E}_\uparrow}{\vekc{E}_\downarrow}
  = \arr{cc}{\rho_{\uparrow\uparrow} & \rho_{\uparrow\downarrow}}
            {\rho_{\downarrow\uparrow} & \rho_{\downarrow\downarrow}}
  \arr{c}{\vek{j}_\uparrow}{\vek{j}_\downarrow} .
\end{equation}
Here, the effective electric fields $\vekc{E}_\sigma$ are the sums
of a spin-independent external electric field plus the gradient of
the local chemical potential, which can be spin-dependent, and
$\vek{j}_\sigma$ is the electric current carried by the electrons
with spin $\sigma$. The spin Coulomb drag gives rise to a spin
trans-resistivity $\rho_{\uparrow\downarrow}$ in equation
(\ref{eq:spindrag}), which is the ratio of the gradient of the
spin-down electrochemical potential to the spin-up current density
when the spin-down current is zero. \citet{dam00} evaluated
$\rho_{\uparrow\downarrow}$ in a generalized random-phase
approximation.

\begin{figure}[t]
  \centerline{\includegraphics[width=0.7\linewidth]{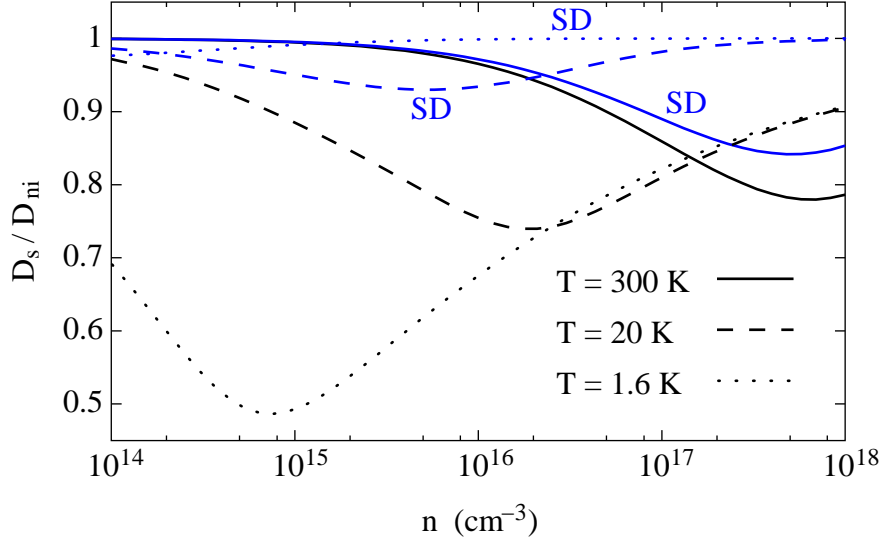}}
  \caption{\label{fig:spindrag}Ratio $D_s/D_\mathrm{ni}$ between the
  spin diffusion coefficient $D_s$ of an interacting electron system
  and the spin diffusion coefficient $D_\mathrm{ni}$ of the
  corresponding noninteracting system, plotted as a function of
  density $n$, and assuming a dielectric constant $\eps=12$
  appropriate for GaAs and mobility $\mu = 3\times
  10^3$~cm$^2$~V$^{-1}$~s$^{-1}$. The curves labeled SD correspond to
  the case in which interactions in $D_s$ are taken into account
  only through the spin Coulomb drag [i.e., the third factor in
  equation (\ref{eq:diffcoul})]. [Adapted from~\citet{dam02}.]}
\end{figure}

\citet{dam01} showed that the Coulomb interaction usually gives rise
to a significant reduction of the spin diffusion coefficient $D_s$
in equation (\ref{eq:drift}) as compared to its value
$D_\mathrm{ni}$ in a noninteracting system. They obtained
\begin{equation}
  \label{eq:diffcoul}
  D_s = \frac{\mu\,\kbolz T}{e} \: \frac{\mathcal{S}}{\mathcal{S}_c} \:
  \frac{1}{1 - \rho_{\uparrow\downarrow}/\rho_\mathrm{D}},
\end{equation}
where $\mu\,\kbolz T/e$ is the diffusion constant of a
noninteracting system in the high-tem\-perature limit (Einstein's
relation), $\mathcal{S}$ is the spin stiffness (i.e., the inverse of
the spin susceptibility), $\mathcal{S}_c = \kbolz T \,
n/(4n_\uparrow n_\downarrow)$ is the Curie spin stiffness of an
ideal classical gas, and $\rho_\mathrm{D} =
m^\ast/(ne^2\tau_\mathrm{D})$ is the Drude resistivity.
Figure~\ref{fig:spindrag} shows the ratio $D_s/D_\mathrm{ni}$ as a
function of density $n$, assuming a dielectric constant $\eps=12$
appropriate for GaAs and mobility $\mu = 3\times
10^3$~cm$^2$~V$^{-1}$~s$^{-1}$. Different line styles correspond to
different temperatures as indicated. The curves labeled SD
correspond to the case in which interactions in $D_s$ are taken into
account only through the spin Coulomb drag [i.e., the third factor
in equation (\ref{eq:diffcoul})].

Figure~\ref{fig:spindrag} shows that the interaction correction is
quite significant and reduces the value of $D_s$. For large
densities or $T \lesssim T_\mathrm{F}$, where $T_\mathrm{F}$ is the
Fermi temperature for density $n$, the dominant effect in the full
calculation stems from the softening of the spin stiffness. On the
other hand, the spin drag contribution dominates at small densities
(the nondegenerate limit $T \gg T_\mathrm{F}$). Note that
$T_\mathrm{F} = 1.6$, 20, and 300~K correspond to $n=7.4\times
10^{15}$, $3.2\times 10^{17}$, and $1.9\times 10^{19}$~cm$^{-3}$,
respectively.

\begin{figure}[tbp]
  \centerline{\includegraphics[width=0.7\linewidth]{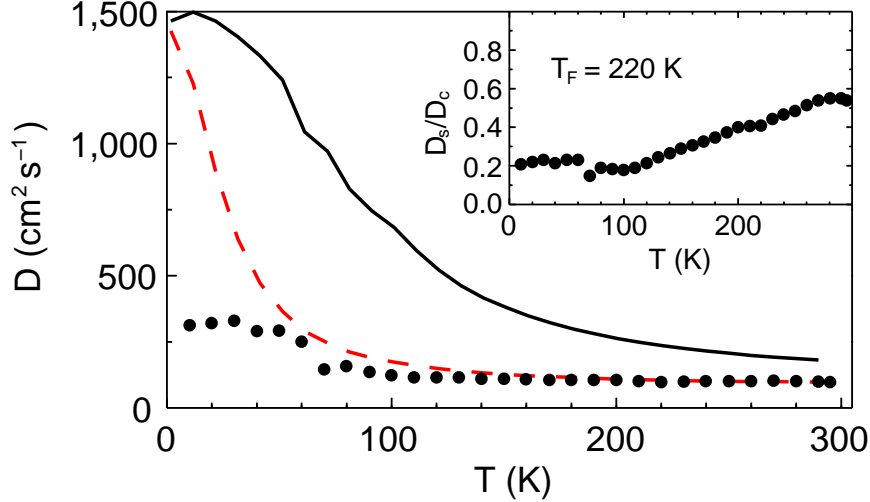}}
  \caption{\label{fig:weber}Measured (circles) and calculated
  (dashed line) spin diffusion coefficient $D_s$ and charge
  diffusion coefficient $D_c$ (solid line) in a 2D electron gas in a
  GaAs/AlGaAs QW. The electron concentration is $n = 4.3 \times
  10^{11}$~cm$^{-2}$, which corresponds to a Fermi temperature
  $T_\mathrm{F} = 220$~K. The inset shows the ratio between the
  measured $D_s$ and $D_c$.
  [Adapted from~\citet{web05}.]}
\end{figure}

The spin Coulomb drag in a 2D electron gas was studied theoretically
by \citet{dam03} and \citet{fle01}, giving results quantitatively
similar to 3D electron systems. It was also observed experimentally
by \citet{web05}. These authors used spin gratings as discussed at
the end of Section~\ref{sec:drift} to measure the spin diffusion
coefficient $D_s$ in a 2D electron gas in a GaAs/AlGaAs QW (circles
in Figure~\ref{fig:weber}). Its value as a function of temperature
is significantly smaller than the charge diffusion coefficient $D_c$
obtained from transport measurements for samples from the same wafer
(solid lines in Figure~\ref{fig:weber}). Yet good agreement is
achieved between the measured $D_s$ and calculations taking into
account the spin Coulomb drag effect [i.e., the last factor in
equation (\ref{eq:diffcoul})], see the dashed line in
Figure~\ref{fig:weber}.

\section{Spin polarization and charge currents}
\label{sec:charge}

\subsection{Spin photocurrents}
\label{sec:photocur}

The optical creation of spin-oriented electrons can give rise to
charge currents, the so-called spin photocurrents, which are
characterized by the fact that these currents reverse their
direction when the radiation helicity is changed from left-handed to
right-handed and vice versa. Spin photocurrents are described by an
axial tensor (or pseudotensor) of second rank. Such tensors play an
important role in the context of gyrotropy, so that systems
permitting nonzero axial second-rank tensors are often denoted
gyrotropic systems. We note that gyrotropy is found neither in
inversion-symmetric systems nor in systems with a zinc blende
structure. The 18 gyrotropic crystal classes are listed, e.g., by
\citet{agr84}.%
\footnote{As certain aspects of gyrotropy require a symmetric
material tensor, the discussion of gyrotropy is often restricted to
those 15 crystal classes that permit a symmetric axial tensor of
second rank \citep{nye57, lanVIIIe}, thus excluding the crystal
classes $C_{3v}$, $C_{4v}$, and $C_{6v}$ (the latter includes
wurtzite materials). Spin photocurrents and the electric generation
of spin discussed below do not require that the corresponding
material tensors are symmetric. Therefore, these effects can be
observed for all 18 crystal classes that permit an axial tensor of
second rank. A general discussion of the symmetry of material
tensors was given, e.g., by \citet{bir74}.}
Semiconductors with a zinc blende (or diamond) structure become
gyrotropic when the symmetry is reduced by means of, e.g., quantum
confinement or uniaxial strain. We note that gyrotropy is also a
required and sufficient condition for the existence of $k$-linear
spin splitting of the energy spectrum of spin-1/2 electron systems.

\begin{figure}[tbp]
  \includegraphics[width=0.35\linewidth]{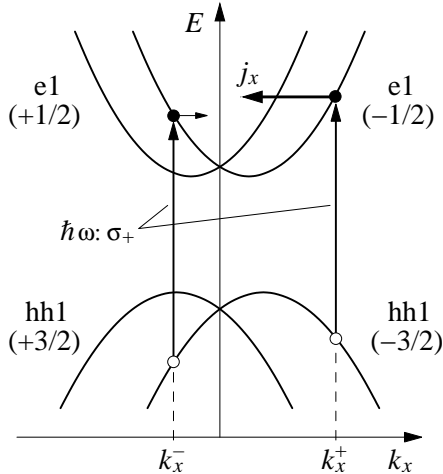}
  \hfill
  \raisebox{2ex}{\begin{minipage}[b]{0.6\linewidth}
    \caption{\label{fig:cpge}Microscopic picture of the circular
    photogalvanic effect [after \citet{gan01}]. $\sigma_+$
    excitation with photon energy $\hbar\omega$ induces optical
    transitions between the valence subband hh1 and the conduction
    subband e1 (vertical arrows). The concurrence of energy
    conservation, spin splitting of the electron and hole states,
    and optical selection rules results in an unbalanced occupation
    of the positive ($k_x^+$) and negative ($k_x^-$) states yielding
    a spin-polarized photocurrent.}
  \end{minipage}}
\end{figure}

Two mechanisms contribute to spin photocurrents in gyrotropic media,
the circular photogalvanic effect and the spin-galvanic effect
\citep{gan03}. The circular photogalvanic effect (CPGE) was
independently predicted by \citet{ivc78} and \citet{bel78}.
Subsequently, this effect was observed in bulk Te by \citet{asn78}.
The mechanism is illustrated in Figure~\ref{fig:cpge}. Excitation
with $\sigma_+$-polarized light induces direct optical transitions
between the valence subband hh1 and the conduction subband e1
(vertical arrows in Figure~\ref{fig:cpge}). For a given photon
energy $\hbar\omega$, the optical selection rules and spin splitting
result in an unbalanced occupation of the positive ($k_x^+$) and
negative ($k_x^-$) states such that the ``centre of mass'' of these
transitions is shifted from $k_x = 0$ to some average value $\langle
k_x \rangle \ne 0$. This wave vector $\langle k_x \rangle$
translates into an average electron velocity $v = \hbar\langle k_x
\rangle / m^\ast$ of the optically oriented electrons, which
corresponds to a spin-polarized charge current, i.e., the current is
carried by electrons with one spin orientation. For interband
transitions in 2D systems, as depicted in Figure~\ref{fig:cpge}, a
detailed theory for the CPGE has been formulated by \citet{gol03}.
Spin photocurrents can also be generated in a similar way by means
of inter-subband and intra-subband transitions \citep{gan01, gan03}.
\citet{she05} and \citet{tar05} have shown that pure spin
photocurrents not accompanied by charge transfer or spin orientation
can be generated by means of absorption of unpolarized light in
low-dimensional semiconductor systems.

Up to now, we have discussed spin photocurrents obtained by means of
one-photon absorption. These currents can also be generated by means
of two-photon excitation \citep{bha00}. In this case, the spin
polarization of the resulting charge currents has been confirmed
directly by measuring the phase-dependent spatial shift of the
circularly polarized photoluminescence \citep{hue03}. Pure spin
photocurrents not accompanied by charge transfer have been generated
through quantum interference of one- and two-photon absorption by
\citet{ste03}.

\begin{figure}[tbp]
  \includegraphics[width=0.35\linewidth]{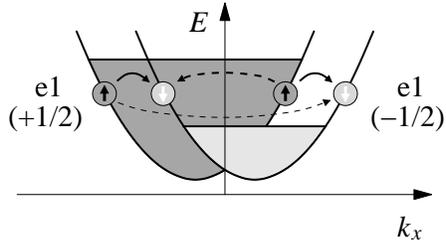}
  \hfill
  \raisebox{2ex}{\begin{minipage}[b]{0.6\linewidth}
    \caption{\label{fig:sge}One-dimensional microscopic picture of
    the spin-galvanic effect [after \citet{gan01}]. If one spin
    subband is preferentially occupied, e.g., by optical excitation,
    asymmetric spin-flip scattering results in a current in the $x$
    direction.}
  \end{minipage}}
\end{figure}

Besides the CPGE, the spin-galvanic effect (SGE) is a second
mechanism that contributes to spin photocurrents \citep{ivc89,
gan02a}. The SGE is caused by asymmetric spin-flip relaxation of
spin-polarized electrons. The mechanism is illustrated in
Figure~\ref{fig:sge}. An unbalanced population of spin-up and
spin-down subbands is generated, e.g., by optical orientation. The
current flow is caused by $\kk$-dependent spin-flip relaxation
processes. Spins oriented in the up direction are scattered along
$k_x$ from the more occupied, e.g., spin-up branch, to the less
filled spin-down branch. Four quantitatively different spin-flip
scattering events exist and are sketched in Figure~\ref{fig:sge} as
bent arrows. The spin-flip scattering rate depends on the values of
the wavevectors of the initial and the final states. Therefore, the
spin-flip transitions marked by solid arrows in Figure~\ref{fig:sge}
have the same rates. They preserve the distribution of carriers in
the branches and, thus, do not yield a current. However, the two
scattering processes shown by dashed arrows are inequivalent and
generate an asymmetric carrier distribution around the branch
minima. This asymmetric population results in a current flow along
the $x$-direction. Within this model of elastic scattering the
current is not spin polarized, since the same number of spin-up and
spin-down electrons move in the same direction with the same
velocity \citep{gan02a}.

Assuming a linear relation between the components $S_\beta$ of the
electrons' averaged spin density and the components $j_\alpha$ of
the resulting spin photocurrent, we get for the SGE
\begin{equation}
  \label{eq:sge}
  j_\alpha = \sum_\beta \mathcal{T}_{\alpha\beta} \, S_\beta
  \qquad \alpha,\beta = x,y,z,
\end{equation}
where $\mathcal{T}_{\alpha\beta}$ is an axial tensor of second
rank.%
\footnote{For many crystal classes permitting nonzero axial tensors
of second rank it is nontheless required by symmetry that certain
components of these tensors must vanish, see, e.g., the discussion
of the experiment of \citet{gan04, gan06a} in
Section~\ref{sec:elspingen}.}
This equation shows clearly that, unlike the case of the CPGE,
optical excitation is not required for the SGE. The CPGE, however,
is always accompanied by the SGE. Formally, this is due to the fact
that both effects are characterized by axial tensors of second rank.
Even in a completely optical experiment, CPGE and SGE can be
distinguished by their different behaviors when the light source is
switched off. Then the circular photogalvanic current decays with
the momentum relaxation time whereas the spin-galvanic current
decays with the spin relaxation time. If spin relaxation is absent,
the spin-galvanic current vanishes \citep{gan03}.

In recent years, detailed experimental and theoretical
investigations of the CPGE and SGE in different systems have been
performed by Ganichev \emph{et al}. This work and related work have
been reviewed by \citet{gan03, gan06}.

\subsection{Electrical generation of a spin polarization}
\label{sec:elspingen}

In general, two possibilities exist for orienting electron spins
with electric currents in a semiconductor. The first one is the spin
Hall effect. For semiconductor systems, this idea was first
discussed by \citet{dya71b}. It yields a spin accumulation at the
\emph{edges} of the sample in the direction perpendicular to the
current. A detailed discussion of the spin Hall effect can be found
\crossref. In gyrotropic media, a second mechanism exists that
yields a spin polarization in the \emph{bulk} of the sample
\citep{aro89, ede90}. We note that equation (\ref{eq:sge}), relating
the given spin orientation $\vek{S}$ with the resulting current
$\vek{j}$, can obviously be inverted, i.e., an electric current
$\vek{j}$ can give rise to a spin density $\vek{S}$ \citep{ivc78}.
As discussed in detail by \citet{aro91}, the different mechanisms
contributing to the spin polarization of electrons induced by a
current $\vek{j}$ can be classified analogously to the different
spin relaxation mechanisms for $\vek{j} = 0$: for $\vek{j} = 0$,
these mechanisms drive the system towards its equilibrium
configuration characterized by equal occupations of the spin states.
For $\vek{j} \ne 0$, on the other hand, the nonequilibrium
configuration is characterized by an unequal filling of the spin
states. Apart from a prefactor $Q$ of order one, the details of
which depend on the scattering mechanisms present in the electron
system, the spin polarization is given by the ratio between the spin
splitting $\hbar\Omega(\kk_\mathcal{E})$ (assumed to be linear in
$\kk$) and the average energy $\bar{E}$ of the involved electrons
\citep{aro91}
\begin{equation}
  \label{eq:spinor}
  \vek{S} = Qn \: \frac{\hbar\vek{\Omega}(\kk_\mathcal{E})}
                 {\bar{E}}.
\end{equation}
Here $\kk_\mathcal{E} = e\vekc{E}\tau_p / \hbar$ is the shift of the
Fermi sphere caused by the electric field $\vekc{E}$, and $\tau_p$
is the momentum relaxation time. In degenerate systems, we have
$\bar{E} = \hbar^2 k_\fermi^2 / (2m^\ast)$. In nondegenerate
systems we have $\bar{E} = (d/2)\, \kbolz T$, where $d$ is the
dimension. Finally, $n=k_\fermi^d/(d\pi)$ is the number density. The
prefactor $Q$ for different scattering mechanisms in $d=2$ and $d=3$
dimensions is given in Table~I of \citet{aro91}.

\begin{figure}[tbp]
  \includegraphics[width=0.40\linewidth]{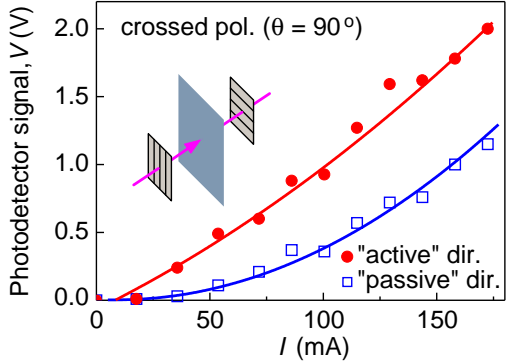}
  \hfill
  \raisebox{2ex}{\begin{minipage}[b]{0.55\linewidth}
    \caption{\label{fig:ganichev}Transmission of a GaAs multi-QW as
    a function of current $I$ in a crossed-polarizer setup (sketched
    in the inset). The sample was grown on a miscut (001) surface so
    that spin orientation is expected for the ``active'' direction
    $[1\overline{1}0]$ but not for the perpendicular ``passive''
    direction. [Adapted from \citet{gan04}.]}
  \end{minipage}}
\end{figure}

The electric-field-induced spin orientation inside a semiconductor
was also studied theoretically by \citet{mag01} in 2D and
\citet{cul05} in 2D and 3D. The effect was first observed
experimentally in bulk Te by \citet{vor79}. More recently, it was
used by \citet{ham99, ham00a} to analyze the spin injection from a
ferromagnetic film into a 2D electron system, see also \citet{mon00,
wee00} and \citet{sil01}. Moreover, the effect was measured in
strained bulk InGaAs by \citet{kat04b} and in 2D GaAs systems by
\citet{sil04} and \citet{gan04, gan06a}. As an example, we want
to discuss the experiment of \citeauthor{gan04} They used a
$p$-type GaAs multi-QW grown on an intentionally miscut (001)
surface (tilted by $5^\circ$ towards the [110] direction). The
symmetry of this system is thus fully characterized by one mirror
plane $(1\overline{1}0)$ (i.e., point group $C_s$), and electric
spin orientation is expected only for a current in the (``active'')
direction $[1\overline{1}0]$ of the 2D plane, but not for the
perpendicular (``passive'') direction. In a transmission measurement
using linearly polarized light, it is then possible to identify the
current-induced spin orientation via a rotation of the polarization
vector of the transmitted light (dichroic absorption and Faraday
rotation) in a crossed polarizer setup, see the inset of
Figure~\ref{fig:ganichev}. For the ``active'' direction
$[1\overline{1}0]$, \citeauthor{gan04} observed a significantly
larger signal in the photodetector than for the ``passive''
direction (Figure~\ref{fig:ganichev}). The nonzero signal for the
``passive'' direction was ascribed to imperfections of the infrared
polarizers and carrier heating by the current, as confirmed by
control experiments.

Finally, we note that \citet{kal90} predicted and observed a
current-induced spin precession in the field
$\vek{\Omega}(\kk_\mathcal{E})$.

\section{Outlook}

We focused here on the fundamental physics underlying the
spin-dependent transport of carriers in semiconductors. These
phenomena have many important and fascinating applications in the
field of spintronics that are discussed \crossref.
Particularly important are various laterally structured systems such
as the Datta-Das spin transistor \citep{dat90} and hybrid devices
combining nonmagnetic semiconductors with semimagnetic and
ferromagnetic materials.

\section*{Acknowledgment}

The author appreciates stimulating discussions with M.~Beck,
D.~Culcer, S.~D.\ Ganichev, and L.~E.\ Golub. Also, he is grateful
to S.~A.\ Crooker for Figure~\ref{fig:crooker}.



\begin{thebibliography}{100}
\providecommand{\natexlab}[1]{#1}

\bibitem[{Agranovich and Ginzburg(1984)}]{agr84}
Agranovich, V.~M. and Ginzburg, V.~L. (1984), \emph{Crystal Optics with Spatial
  Dispersion, and Excitons}, Springer, Berlin, 2nd ed.

\bibitem[{Aronov and Lyanda-Geller(1989)}]{aro89}
Aronov, A.~G. and Lyanda-Geller, Y.~B. (1989), Nuclear electric resonance and
  orientation of carrier spins by an electric field, \emph{JETP Lett.}
  \textbf{50}, 431--434.

\bibitem[{Aronov \emph{et~al.}(1991)Aronov, Lyanda-Geller, and Pikus}]{aro91}
Aronov, A.~G., Lyanda-Geller, Y.~B., and Pikus, G.~E. (1991), Spin polarization
  of electrons by an electric current, \emph{Sov. Phys.--JETP} \textbf{73},
  537--541.

\bibitem[{Asnin \emph{et~al.}(1978)Asnin, Bakun, Danishevski\u{\i}, Ivchenko,
  Pikus, and Rogachev}]{asn78}
Asnin, V.~M., Bakun, A.~A., Danishevski\u{\i}, A.~M., Ivchenko, E.~L., Pikus,
  G.~E., and Rogachev, A.~A. (1978), Observation of a photo-emf that depends on
  the sign of the circular polarization of the light, \emph{JETP Lett.}
  \textbf{28}, 74--77.

\bibitem[{Beck \emph{et~al.}(2005)Beck, Metzner, Malzer, and D\"ohler}]{bec05}
Beck, M., Metzner, C., Malzer, S., and D\"ohler, G.~H. (2005), Spin lifetimes
  and strain-controlled spin precession of drifting electrons in zinc blende
  type semiconductors, cond-mat/0504668.

\bibitem[{Belinicher(1978)}]{bel78}
Belinicher, V. (1978), Space-oscillating photocurrent in crystals without
  symmetry center, \emph{Phys. Lett.} \textbf{66A}, 213--214.

\bibitem[{Bhat and Sipe(2000)}]{bha00}
Bhat, R. D.~R. and Sipe, J.~E. (2000), Optically injected spin currents in
  semiconductors, \emph{Phys.\ Rev.\ Lett.} \textbf{85}, 5432--5435.

\bibitem[{Bir and Pikus(1974)}]{bir74}
Bir, G.~L. and Pikus, G.~E. (1974), \emph{Symmetry and Strain-Induced Effects
  in Semiconductors}, Wiley, New York.

\bibitem[{Bleibaum(2006)}]{ble06}
Bleibaum, O. (2006), Spin diffusion equations for systems with {Rashba}
  spin-orbit interaction in an electric field, \emph{Phys.\ Rev.~B}
  \textbf{73}, 035322.

\bibitem[{Braun and R\"ossler(1985)}]{bra85}
Braun, M. and R\"ossler, U. (1985), Magneto-optic transitions and
  non-parabolicity parameters in the conduction band of semiconductors,
  \emph{J.\ Phys.~C: Solid State Phys.} \textbf{18}, 3365--3377.

\bibitem[{Bychkov and Rashba(1984)}]{byc84}
Bychkov, Y.~A. and Rashba, E.~I. (1984), Oscillatory effects and the magnetic
  susceptibility of carriers in inversion layers, \emph{J.\ Phys.~C: Solid
  State Phys.} \textbf{17}, 6039--6045.

\bibitem[{Cameron \emph{et~al.}(1996)Cameron, Riblet, and Miller}]{cam96}
Cameron, A.~R., Riblet, P., and Miller, A. (1996), Spin gratings and the
  measurement of electron drift mobility in multiple quantum well
  semiconductors, \emph{Phys.\ Rev.\ Lett.} \textbf{76}, 4793--4796.

\bibitem[{Crooker and Smith(2005)}]{cro05}
Crooker, S.~A. and Smith, D.~L. (2005), Imaging spin flows in semiconductors
  subject to electric, magnetic, and strain fields, \emph{Phys.\ Rev.\ Lett.}
  \textbf{94}, 236601.

\bibitem[{Culcer \emph{et~al.}(2005)Culcer, Yao, MacDonald, and Niu}]{cul05}
Culcer, D., Yao, Y., MacDonald, A.~H., and Niu, Q. (2005), Electrical
  generation of spin in crystals with reduced symmetry, \emph{Phys.\ Rev.~B}
  \textbf{72}, 045215.

\bibitem[{{D'Amico} and Vignale(2000)}]{dam00}
{D'Amico}, I. and Vignale, G. (2000), Theory of spin {Coulomb} drag in
  spin-polarized transport, \emph{Phys.\ Rev.~B} \textbf{62}, 4853--4857.

\bibitem[{{D'Amico} and Vignale(2001)}]{dam01}
{D'Amico}, I. and Vignale, G. (2001), Spin diffusion in doped semiconductors:
  The role of {Coulomb} interactions, \emph{Europhys.\ Lett.} \textbf{55},
  566--572.

\bibitem[{{D'Amico} and Vignale(2002)}]{dam02}
{D'Amico}, I. and Vignale, G. (2002), Coulomb interaction effects in
  spin-polarized transport, \emph{Phys.\ Rev.~B} \textbf{65}, 085109.

\bibitem[{{D'Amico} and Vignale(2003)}]{dam03}
{D'Amico}, I. and Vignale, G. (2003), Spin {Coulomb} drag in the
  two-dimen\-sion\-al electron liquid, \emph{Phys.\ Rev.~B} \textbf{68},
  045307.

\bibitem[{Dareys \emph{et~al.}(1993)Dareys, Marie, Amand, Barrau, Shekun,
  Razdobreev, and Planel}]{dar93a}
Dareys, B., Marie, X., Amand, T., Barrau, J., Shekun, Y., Razdobreev, I., and
  Planel, R. (1993), Spin dynamics of exciton states in {GaAs}/{AlGaAs}
  multiple quantum wells, \emph{Superlatt. Microstruct.} \textbf{13}, 353--358.

\bibitem[{Datta and Das(1990)}]{dat90}
Datta, S. and Das, B. (1990), Electronic analog of the electro-optic modulator,
  \emph{Appl.\ Phys.\ Lett.} \textbf{56}, 665--667.

\bibitem[{{de Andrada e Silva}(1992)}]{and92}
{de Andrada e Silva}, E.~A. (1992), Conduction-subband anisotropic spin
  splitting in {III-V} semiconductor heterojunctions, \emph{Phys.\ Rev.~B}
  \textbf{46}, 1921--1924.

\bibitem[{Dresselhaus(1955)}]{dre55a}
Dresselhaus, G. (1955), Spin-orbit coupling effects in zinc blende structures,
  \emph{Phys.\ Rev.} \textbf{100}, 580--586.

\bibitem[{Dumke(1957)}]{dum57}
Dumke, W.~P. (1957), Spontaneous radiative recombination in semiconductors,
  \emph{Phys.\ Rev.} \textbf{105}, 139--144.

\bibitem[{D'yakonov \emph{et~al.}(1986)D'yakonov, Marushchak, Perel', and
  Titkov}]{dya86a}
D'yakonov, M.~I., Marushchak, V.~A., Perel', V.~I., and Titkov, A.~N. (1986),
  The effect of strain on the spin relaxation of conduction electrons in
  {III-V} semiconductors, \emph{Sov. Phys.--JETP} \textbf{63}, 655--661.

\bibitem[{D'yakonov and Perel'(1971{\natexlab{a}})}]{dya71a}
D'yakonov, M.~I. and Perel', V.~I. (1971{\natexlab{a}}), Feasibility of optical
  orientation of equilibrium electrons in semiconductors, \emph{JETP Lett.}
  \textbf{13}, 144--146.

\bibitem[{D'yakonov and Perel'(1971{\natexlab{b}})}]{dya71b}
D'yakonov, M.~I. and Perel', V.~I. (1971{\natexlab{b}}), Possibility of
  orienting electron spins with current, \emph{JETP Lett.} \textbf{13},
  467--469.

\bibitem[{D'yakonov and Perel'(1971{\natexlab{c}})}]{dya71}
D'yakonov, M.~I. and Perel', V.~I. (1971{\natexlab{c}}), Spin orientation of
  electrons associated with the interband absorption of light in
  semiconductors, \emph{Sov. Phys.--JETP} \textbf{33}, 1053--1059.

\bibitem[{D'yakonov and Perel'(1976)}]{dya76}
D'yakonov, M.~I. and Perel', V.~I. (1976), Theory of the {Hanle} effect in
  optical orientation of electrons in $n$-type semiconductors, \emph{Sov.
  Phys.--Semicond.} \textbf{10}, 350--353.

\bibitem[{Dyakonov and Perel(1984)}]{dya84}
Dyakonov, M.~I. and Perel, V.~I. (1984), Theory of optical spin orientation of
  electrons and nuclei in semiconductors, in  \cite{mei84}, pp. 11--71.

\bibitem[{Dzhioev \emph{et~al.}(1997)Dzhioev, Zakharchenya, Korenev, and
  Stepanova}]{dzh97}
Dzhioev, R.~I., Zakharchenya, B.~P., Korenev, V.~L., and Stepanova, M.~N.
  (1997), Spin diffusion of optically oriented electrons and photon entrainment
  in $n$-gallium arsenide, \emph{Phys. Solid State} \textbf{39}, 1765--1768.

\bibitem[{Edelstein(1990)}]{ede90}
Edelstein, V.~M. (1990), Spin polarization of conduction electrons induced by
  electron current in two-dimen\-sion\-al asymmetric electron systems,
  \emph{Solid State Commun.} \textbf{73}, 233--235.

\bibitem[{Fiederling \emph{et~al.}(1999)Fiederling, Keim, Reuscher, Ossau,
  Schmidt, Waag, and Molenkamp}]{fie99}
Fiederling, R., Keim, M., Reuscher, G., Ossau, W., Schmidt, G., Waag, A., and
  Molenkamp, L.~W. (1999), Injection and detection of a spin-polarized current
  in a light-emitting diode, \emph{Nature} \textbf{402}, 787--790.

\bibitem[{Flatt\'e and Byers(2000)}]{fla00}
Flatt\'e, M.~E. and Byers, J.~M. (2000), Spin diffusion in semiconductors,
  \emph{Phys.\ Rev.\ Lett.} \textbf{84}, 4220--4223.

\bibitem[{Flensberg \emph{et~al.}(2001)Flensberg, Jensen, and
  Mortensen}]{fle01}
Flensberg, K., Jensen, T.~S., and Mortensen, N.~A. (2001), Diffusion equation
  and spin drag in spin-polarized transport, \emph{Phys.\ Rev.~B} \textbf{64},
  245308.

\bibitem[{Freeman \emph{et~al.}(1990)Freeman, Awschalom, and Hong}]{fre90}
Freeman, M.~R., Awschalom, D.~D., and Hong, J.~M. (1990), Picosecond
  photoluminescence excitation spectroscopy of {GaAs}/{AlGaAs} quantum wells,
  \emph{Appl.\ Phys.\ Lett.} \textbf{57}, 704--706.

\bibitem[{Ganichev and Prettl(2006)}]{gan06}
Ganichev, S. and Prettl, W. (2006), \emph{Intense Terahertz Excitation of
  Semiconductors}, Oxford University Press, New York.

\bibitem[{Ganichev \emph{et~al.}(2004)Ganichev, Danilov, Schneider, Bel'kov,
  Golub, Wegscheider, Weiss, and Prettl}]{gan04}
Ganichev, S.~D., Danilov, S.~N., Schneider, P., Bel'kov, V.~V., Golub, L.~E.,
  Wegscheider, W., Weiss, D., and Prettl, W. (2004), Can an electric current
  orient spins in quantum wells?, cond-mat/0403641.

\bibitem[{Ganichev \emph{et~al.}(2006)Ganichev, Danilov, Schneider, Bel'kov,
  Golub, Wegscheider, Weiss, and Prettl}]{gan06a}
Ganichev, S.~D., Danilov, S.~N., Schneider, P., Bel'kov, V.~V., Golub, L.~E.,
  Wegscheider, W., Weiss, D., and Prettl, W. (2006), Electric current-induced
  spin orientation in quantum well structures, \emph{J. Magn. Magn. Mater.}
  \textbf{300}, 127--131.

\bibitem[{Ganichev \emph{et~al.}(2002)Ganichev, Ivchenko, Bel'kov, Tarasenko,
  Sollinger, Weiss, Wegscheider, and Prettl}]{gan02a}
Ganichev, S.~D., Ivchenko, E.~L., Bel'kov, V.~V., Tarasenko, S.~A., Sollinger,
  M., Weiss, D., Wegscheider, W., and Prettl, W. (2002), Spin-galvanic effect,
  \emph{Nature} \textbf{417}, 153--156.

\bibitem[{Ganichev \emph{et~al.}(2001)Ganichev, Ivchenko, Danilov, Eroms,
  Wegscheider, Weiss, and Prettl}]{gan01}
Ganichev, S.~D., Ivchenko, E.~L., Danilov, S.~N., Eroms, J., Wegscheider, W.,
  Weiss, D., and Prettl, W. (2001), Conversion of spin into directed electric
  current in quantum wells, \emph{Phys.\ Rev.\ Lett.} \textbf{86}, 4358--4361.

\bibitem[{Ganichev and Prettl(2003)}]{gan03}
Ganichev, S.~D. and Prettl, W. (2003), Spin photocurrents in quantum wells,
  \emph{J.\ Phys.: Condens.\ Mat.} \textbf{15}, R935--R983.

\bibitem[{Garbuzov \emph{et~al.}(1976)Garbuzov, Merkulov, Novikov, and
  Fleisher}]{gar76}
Garbuzov, D.~Z., Merkulov, I.~A., Novikov, V.~A., and Fleisher, V.~G. (1976),
  Diffusion of `spin-labelled' electrons in a double heterostructure,
  \emph{Sov. Phys.--Semicond.} \textbf{10}, 552--555.

\bibitem[{Gioev \emph{et~al.}(1994)Gioev, Zakharchenya, Kavokin, and
  Pak}]{gio94}
Gioev, R.~I., Zakharchenya, B.~P., Kavokin, K.~V., and Pak, P.~E. (1994), Study
  of diffusional and radiative electron transport in $p$-{GaAs} by the optical
  orientation method, \emph{Phys. Solid State} \textbf{36}, 1501--1506.

\bibitem[{Golub(2003)}]{gol03}
Golub, L.~E. (2003), Spin-splitting-induced photogalvanic effect in quantum
  wells, \emph{Phys.\ Rev.~B} \textbf{67}, 235320.

\bibitem[{Gramila \emph{et~al.}(1991)Gramila, Eisenstein, MacDonald, Pfeiffer,
  and West}]{gra91}
Gramila, T.~J., Eisenstein, J.~P., MacDonald, A.~H., Pfeiffer, L.~N., and West,
  K.~W. (1991), Mutual friction between parallel two-dimen\-sion\-al electron
  systems, \emph{Phys.\ Rev.\ Lett.} \textbf{66}, 1216--1219.

\bibitem[{H\"agele \emph{et~al.}(1998)H\"agele, Oestreich, R\"uhle, Nestle, and
  Eberl}]{hae98}
H\"agele, D., Oestreich, M., R\"uhle, W.~W., Nestle, N., and Eberl, K. (1998),
  Spin transport in {GaAs}, \emph{Appl.\ Phys.\ Lett.} \textbf{73}, 1580--1582.

\bibitem[{Hammar \emph{et~al.}(1999)Hammar, Bennett, Yang, and Johnson}]{ham99}
Hammar, P.~R., Bennett, B.~R., Yang, M.~J., and Johnson, M. (1999), Observation
  of spin injection at a ferromagnet-semiconductor interface, \emph{Phys.\
  Rev.\ Lett.} \textbf{83}, 203--206.

\bibitem[{Hammar \emph{et~al.}(2000)Hammar, Bennett, Yang, and
  Johnson}]{ham00a}
Hammar, P.~R., Bennett, B.~R., Yang, M.~J., and Johnson, M. (2000), A reply to
  the comment by {F. G. Monzon, H. X. Tang, and M. L. Roukes}, and also to the
  comment by {B. J. van Wees}., \emph{Phys.\ Rev.\ Lett.} \textbf{84},
  5024--5025.

\bibitem[{Howlett and Zukotynski(1977)}]{how77}
Howlett, W. and Zukotynski, S. (1977), Effect of deformation on the conduction
  band of {III-V} semiconductors, \emph{Phys.\ Rev.~B} \textbf{16}, 3688--3693.

\bibitem[{Hru\v{s}ka \emph{et~al.}(2006)Hru\v{s}ka, \v{S}. Kos, Crooker,
  Saxena, and Smith}]{hru06}
Hru\v{s}ka, M., \v{S}. Kos, Crooker, S.~A., Saxena, A., and Smith, D.~L.
  (2006), Effects of strain, electric, and magnetic fields on lateral
  electron-spin transport in semiconductor epilayers, \emph{Phys.\ Rev.~B}
  \textbf{73}, 075306.

\bibitem[{H\"ubner \emph{et~al.}(2003)H\"ubner, R\"uhle, Klude, Hommel, Bhat,
  Sipe, and van Driel}]{hue03}
H\"ubner, J., R\"uhle, W.~W., Klude, M., Hommel, D., Bhat, R. D.~R., Sipe,
  J.~E., and van Driel, H.~M. (2003), Direct observation of optically injected
  spin-polarized currents in semiconductors, \emph{Phys.\ Rev.\ Lett.}
  \textbf{90}, 216601.

\bibitem[{Ivchenko \emph{et~al.}(1989)Ivchenko, Lyanda-Geller, and
  Pikus}]{ivc89}
Ivchenko, E.~L., Lyanda-Geller, Y.~B., and Pikus, G.~E. (1989), Photocurrent in
  structures with quantum wells with an optical orientation of free carriers,
  \emph{JETP Lett.} \textbf{50}, 175--177.

\bibitem[{Ivchenko and Pikus(1978)}]{ivc78}
Ivchenko, E.~L. and Pikus, G.~E. (1978), New photogalvanic effect in gyrotropic
  crystals, \emph{JETP Lett.} \textbf{27}, 604--608.

\bibitem[{Kalevich and Korenev(1990)}]{kal90}
Kalevich, V.~K. and Korenev, V.~L. (1990), Effect of electric field on the
  optical orientation of {2D} electrons, \emph{JETP Lett.} \textbf{52},
  230--235.

\bibitem[{Kato \emph{et~al.}(2004{\natexlab{a}})Kato, Myers, Gossard, and
  Awschalom}]{kat04a}
Kato, Y., Myers, R.~C., Gossard, A.~C., and Awschalom, D.~D.
  (2004{\natexlab{a}}), Coherent spin manipulation without magnetic fields in
  strained semiconductors, \emph{Nature} \textbf{427}, 50--53.

\bibitem[{Kato \emph{et~al.}(2004{\natexlab{b}})Kato, Myers, Gossard, and
  Awschalom}]{kat04b}
Kato, Y.~K., Myers, R.~C., Gossard, A.~C., and Awschalom, D.~D.
  (2004{\natexlab{b}}), Current-induced spin polarization in strained
  semiconductors, \emph{Phys.\ Rev.\ Lett.} \textbf{93}, 176601.

\bibitem[{Kavokin(2002)}]{kav02}
Kavokin, K.~V. (2002), Optical manifestations of electron spin transport and
  relaxation in semiconductors, \emph{Phys.\ Status Solidi~A} \textbf{190},
  221--227.

\bibitem[{Kikkawa and Awschalom(1999)}]{kik99}
Kikkawa, J.~M. and Awschalom, D.~D. (1999), Lateral drag of spin coherence in
  gallium arsenide, \emph{Nature} \textbf{397}, 139--141.

\bibitem[{Lampel(1968)}]{lam68}
Lampel, G. (1968), Nuclear dynamic polarization by optical electronic
  saturation and optical pumping in semiconductors, \emph{Phys.\ Rev.\ Lett.}
  \textbf{20}, 491--493.

\bibitem[{Landau and Lifshitz(1984)}]{lanVIIIe}
Landau, L.~D. and Lifshitz, E.~M. (1984), \emph{Electrodynamics of Continuous
  Media}, Pergamon, Oxford, 2nd ed.

\bibitem[{Leggett(1970)}]{leg70}
Leggett, A.~J. (1970), Spin diffusion and spin echoes in liquid {$^3$He} at low
  temperature, \emph{J.\ Phys.~C: Solid State Phys.} \textbf{3}, 448--459.

\bibitem[{Leggett and Rice(1968)}]{leg68}
Leggett, A.~J. and Rice, M.~J. (1968), Spin echoes in liquid {He$^3$} and
  mixtures: A predicted new effect, \emph{Phys.\ Rev.\ Lett.} \textbf{20},
  586--589.

\bibitem[{Magarill \emph{et~al.}(2001)Magarill, Chaplik, and \'Entin}]{mag01}
Magarill, L.~I., Chaplik, A.~V., and \'Entin, M.~V. (2001), Spin response of
  {2D} electrons to a lateral electric field, \emph{Semiconductors}
  \textbf{35}, 1081--1087.

\bibitem[{Maialle \emph{et~al.}(1993)Maialle, {de Andrada e Silva}, and
  Sham}]{mai93}
Maialle, M.~Z., {de Andrada e Silva}, E.~A., and Sham, L.~J. (1993), Exciton
  spin dynamics in quantum wells, \emph{Phys.\ Rev.~B} \textbf{47},
  15776--15788.

\bibitem[{Malajovich \emph{et~al.}(2000)Malajovich, Kikkawa, Awschalom, Berry,
  and Samarth}]{mal00}
Malajovich, I., Kikkawa, J.~M., Awschalom, D.~D., Berry, J.~J., and Samarth, N.
  (2000), Coherent transfer of spin through a semiconductor heterointerface,
  \emph{Phys.\ Rev.\ Lett.} \textbf{84}, 1015--1018.

\bibitem[{Masselink \emph{et~al.}(1984)Masselink, Sun, Fischer, Drummond,
  Chang, Klein, and Morko\c{c}}]{mas84}
Masselink, W.~T., Sun, Y.~L., Fischer, R., Drummond, T.~J., Chang, Y.~C.,
  Klein, M.~V., and Morko\c{c}, H. (1984), Enhanced luminescence from
  {AlGaAs}/{GaAs} single quantum well structures through improved interfaces,
  \emph{J.\ Vac.\ Sci.\ Technol.~B} \textbf{2}, 117--122.

\bibitem[{Meier and Zakharchenya(1984)}]{mei84}
Meier, F. and Zakharchenya, B.~P. (Eds.) (1984), \emph{Optical Orientation},
  Elsevier, Amsterdam.

\bibitem[{Monzon \emph{et~al.}(2000)Monzon, Tang, and Roukes}]{mon00}
Monzon, F.~G., Tang, H.~X., and Roukes, M.~L. (2000), Magnetoelectronic
  phenomena at a ferromagnet-semiconductor interface, \emph{Phys.\ Rev.\ Lett.}
  \textbf{84}, 5022.

\bibitem[{{Mu\~{n}oz} \emph{et~al.}(1995){Mu\~{n}oz}, P\'{e}rez, {Vi\~{n}a},
  and Ploog}]{mun95}
{Mu\~{n}oz}, L., P\'{e}rez, E., {Vi\~{n}a}, L., and Ploog, K. (1995), Spin
  relaxation in intrinsic {GaAs} quantum wells: Influence of excitonic
  localization, \emph{Phys.\ Rev.~B} \textbf{51}, 4247--4257.

\bibitem[{Nitta \emph{et~al.}(1997)Nitta, Akazaki, Takayanagi, and
  Enoki}]{nit97}
Nitta, J., Akazaki, T., Takayanagi, H., and Enoki, T. (1997), Gate control of
  spin-orbit interaction in an inverted
  {In$_{0.53}$Ga$_{0.47}$As/In$_{0.52}$Al$_{0.48}$As} heterostructure,
  \emph{Phys.\ Rev.\ Lett.} \textbf{78}, 1335--1338.

\bibitem[{Nye(1957)}]{nye57}
Nye, J.~F. (1957), \emph{Physical Properties of Crystals}, Oxford University
  Press, Oxford.

\bibitem[{Ohkawa and Uemura(1974)}]{ohk74}
Ohkawa, F.~J. and Uemura, Y. (1974), Quantized surface states of narrow-gap
  semiconductors, \emph{J.~Phys.\ Soc.\ Jpn.} \textbf{37}, 1325--1333.

\bibitem[{Ohno \emph{et~al.}(1999)Ohno, Young, Beschoten, Matsukura, Ohno, and
  Awschalom}]{ohn99a}
Ohno, Y., Young, D.~K., Beschoten, B., Matsukura, F., Ohno, H., and Awschalom,
  D.~D. (1999), Electrical spin injection in a ferromagnetic semiconductor
  heterostructure, \emph{Nature} \textbf{427}, 790--792.

\bibitem[{Pfalz \emph{et~al.}(2005)Pfalz, Winkler, Nowitzki, Reuter, Wieck,
  H\"agele, and Oestreich}]{pfa05}
Pfalz, S., Winkler, R., Nowitzki, T., Reuter, D., Wieck, A.~D., H\"agele, D.,
  and Oestreich, M. (2005), Optical orientation of electron spins in {GaAs}
  quantum wells, \emph{Phys.\ Rev.~B} \textbf{71}, 165305.

\bibitem[{Pikus \emph{et~al.}(1988)Pikus, Marushchak, and Titkov}]{pik88}
Pikus, G.~E., Marushchak, V.~A., and Titkov, A.~N. (1988), Spin splitting of
  energy bands and spin relaxation of carriers in cubic {III}-{V} crystals
  (review), \emph{Sov. Phys.--Semicond.} \textbf{22}, 115--124.

\bibitem[{Pikus and Titkov(1984)}]{pik84}
Pikus, G.~E. and Titkov, A.~N. (1984), Spin relaxation under optical
  orientation in semiconductors, in  \cite{mei84}, pp. 73--131.

\bibitem[{Qi and Zhang(2003)}]{qi03}
Qi, Y. and Zhang, S. (2003), Spin diffusion at finite electric and magnetic
  fields, \emph{Phys.\ Rev.~B} \textbf{67}, 052407.

\bibitem[{Ranvaud \emph{et~al.}(1979)Ranvaud, Trebin, R\"ossler, and
  Pollak}]{ran79}
Ranvaud, R., Trebin, H.-R., R\"ossler, U., and Pollak, F.~H. (1979), Quantum
  resonances in the valence bands of zinc-blende semiconductors. {II}.
  {R}esults for $p$-{InSb} under uniaxial stress, \emph{Phys.\ Rev.~B}
  \textbf{20}, 701--715.

\bibitem[{Rashba(1960)}]{ras60}
Rashba, E.~I. (1960), Properties of semiconductors with an extremum loop: {I.}
  {Cyclotron} and combinational resonance in a magnetic field perpendicular to
  the plane of the loop, \emph{Sov. Phys.--Solid State} \textbf{2}, 1109--1122.

\bibitem[{Riechert \emph{et~al.}(1984)Riechert, Alvarado, Titkov, and
  Safarov}]{rie84}
Riechert, H., Alvarado, S.~F., Titkov, A.~N., and Safarov, V.~I. (1984),
  Precession of the spin polarization of photoexcited conduction electrons in
  the band-bending region of {GaAs}(001), \emph{Phys.\ Rev.\ Lett.}
  \textbf{52}, 2297--2300.

\bibitem[{Rojo(1999)}]{roj99}
Rojo, A.~G. (1999), Electron-drag effects in coupled electron systems,
  \emph{J.\ Phys.: Condens.\ Mat.} \textbf{11}, R31--R52.

\bibitem[{R\"ossler and Kainz(2002)}]{ros02}
R\"ossler, U. and Kainz, J. (2002), Microscopic interface asymmetry and
  spin-splitting of electron subbands in semiconductor quantum structures,
  \emph{Solid State Commun.} \textbf{121}, 313--316.

\bibitem[{Roussignol \emph{et~al.}(1992)Roussignol, Rolland, Ferreira,
  Delalande, Bastard, Vinattieri, Carraresi, Colocci, and Etienne}]{rou92}
Roussignol, P., Rolland, P., Ferreira, R., Delalande, C., Bastard, G.,
  Vinattieri, A., Carraresi, L., Colocci, M., and Etienne, B. (1992),
  Time-resolved spin-polarization spectroscopy in {GaAs}/{AlGaAs} quantum
  wells, \emph{Surf.\ Sci.} \textbf{267}, 360--364.

\bibitem[{Seiler \emph{et~al.}(1977)Seiler, Bajaj, and Stephens}]{sei77}
Seiler, D.~G., Bajaj, B.~D., and Stephens, A.~E. (1977), Inversion-asymmetry
  splitting of the conduction band in {InSb}, \emph{Phys.\ Rev.~B} \textbf{16},
  2822--2833.

\bibitem[{Sherman \emph{et~al.}(2005)Sherman, Najmaie, and Sipe}]{she05}
Sherman, E.~Y., Najmaie, A., and Sipe, J.~E. (2005), Spin current injection by
  intersubband transitions in quantum wells, \emph{Appl.\ Phys.\ Lett.}
  \textbf{86}, 122103.

\bibitem[{Silov \emph{et~al.}(2004)Silov, Blajnov, Wolter, Hey, Ploog, and
  Averkiev}]{sil04}
Silov, A.~Y., Blajnov, P.~A., Wolter, J.~H., Hey, R., Ploog, K.~H., and
  Averkiev, N.~S. (2004), Current-induced spin polarization at a single
  heterojunction, \emph{Appl.\ Phys.\ Lett.} \textbf{85}, 5929--5931.

\bibitem[{Silsbee(2001)}]{sil01}
Silsbee, R.~H. (2001), Theory of the detection of current-induced spin
  polarization in a two-dimensional electron gas, \emph{Phys.\ Rev.~B}
  \textbf{63}, 155305.

\bibitem[{Smith(1978)}]{smi78}
Smith, R.~A. (1978), \emph{Semiconductors}, Cambridge University Press,
  Cambridge, 2nd ed.

\bibitem[{Stevens \emph{et~al.}(2003)Stevens, Smirl, Bhat, Najmaie, Sipe, and
  van Driel}]{ste03}
Stevens, M.~J., Smirl, A.~L., Bhat, R. D.~R., Najmaie, A., Sipe, J.~E., and van
  Driel, H.~M. (2003), Quantum interference control of ballistic pure spin
  currents in semiconductors, \emph{Phys.\ Rev.\ Lett.} \textbf{90}, 136603.

\bibitem[{Takahashi \emph{et~al.}(1999)Takahashi, Shizume, and
  Masuhara}]{tak99}
Takahashi, Y., Shizume, K., and Masuhara, N. (1999), Spin diffusion in a
  two-dimen\-sion\-al electron gas, \emph{Phys.\ Rev.~B} \textbf{60},
  4856--4865.

\bibitem[{Tarasenko and Ivchenko(2005)}]{tar05}
Tarasenko, S.~A. and Ivchenko, E.~L. (2005), Pure spin photocurrents in
  low-dimensional structures, \emph{JETP Lett.} \textbf{81}, 231--235.

\bibitem[{Trebin \emph{et~al.}(1979)Trebin, R\"ossler, and Ranvaud}]{tre79}
Trebin, H.-R., R\"ossler, U., and Ranvaud, R. (1979), Quantum resonances in the
  valence bands of zinc-blende semiconductors. {I}. {Theoretical} aspects,
  \emph{Phys.\ Rev.~B} \textbf{20}, 686--700.

\bibitem[{Twardowski and Hermann(1987)}]{twa87}
Twardowski, A. and Hermann, C. (1987), Variational calculation of polarization
  of quantum-well photoluminescence, \emph{Phys.\ Rev.~B} \textbf{35},
  8144--8153.

\bibitem[{Uenoyama and Sham(1990)}]{uen90a}
Uenoyama, T. and Sham, L.~J. (1990), Carrier relaxation and luminescence
  polarization in quantum wells, \emph{Phys.\ Rev.~B} \textbf{42}, 7114--7123.

\bibitem[{van Wees(2000)}]{wee00}
van Wees, B.~J. (2000), Comment on ``{Observation} of spin injection at a
  ferromagnet-semiconductor interface'', \emph{Phys.\ Rev.\ Lett.} \textbf{84},
  5023.

\bibitem[{Vorob'ev \emph{et~al.}(1979)Vorob'ev, Ivchenko, Pikus,
  Farbshte\u{\i}n, Shalygin, and Shturbin}]{vor79}
Vorob'ev, L.~E., Ivchenko, E.~L., Pikus, G.~E., Farbshte\u{\i}n, I.~I.,
  Shalygin, V.~A., and Shturbin, A.~V. (1979), Optical activity in tellurium
  induced by a current, \emph{JETP Lett.} \textbf{29}, 441--445.

\bibitem[{Weber \emph{et~al.}(2005)Weber, Gedik, Moore, Orenstein, Stephens,
  and Awschalom}]{web05}
Weber, C.~P., Gedik, N., Moore, J.~E., Orenstein, J., Stephens, J., and
  Awschalom, D.~D. (2005), Observation of spin {Coulomb} drag in a
  two-dimen\-sion\-al electron gas, \emph{Nature} \textbf{437}, 1330--1333.

\bibitem[{Weisbuch \emph{et~al.}(1981)Weisbuch, Miller, Dingle, Gossard, and
  Wiegmann}]{wei81a}
Weisbuch, C., Miller, R.~C., Dingle, R., Gossard, A.~C., and Wiegmann, W.
  (1981), Intrinsic radiative recombination from quantum states in
  {GaAs}-{AlGaAs} multi-quantum well structures, \emph{Solid State Commun.}
  \textbf{37}, 219--222.

\bibitem[{Winkler(2003)}]{win03}
Winkler, R. (2003), \emph{Spin-Orbit Coupling Effects in Two-Dimen\-sion\-al
  Electron and Hole Systems}, Springer, Berlin.

\bibitem[{Yu and Flatt\'e(2002)}]{yu02}
Yu, Z.~G. and Flatt\'e, M.~E. (2002), Electric-field dependent spin diffusion
  and spin injection into semiconductors, \emph{Phys.\ Rev.~B} \textbf{66},
  201202.

\end{thebibliography}
\end{document}